\documentclass{article}
\usepackage{spconf,amsmath,epsfig,bm,amssymb,amsthm}
\usepackage{psfrag}

\def\L{{\cal L}}

\title{A WIDEBAND SPECTRUM SENSING METHOD FOR COGNITIVE RADIO USING SUB-NYQUIST SAMPLING }

\name{Moslem Rashidi, Kasra Haghighi, Arash Owrang, Mats Viberg \thanks{}}
\address{Department of Signal \& Systems, Chalmers University of Technology}

\begin{document}
\maketitle
\begin{abstract}
\label{sec:abs}
Spectrum sensing is a fundamental component in cognitive radio. A major challenge in this area is the requirement of a high sampling rate in the sensing of a wideband signal. In this paper a wideband spectrum sensing model is presented that utilizes a sub-Nyquist sampling scheme to bring substantial savings in terms of the sampling rate. The correlation matrix of a finite number of noisy samples is computed and used by a subspace estimator to detect the occupied and vacant channels of the spectrum. In contrast with common methods, the proposed method does not need the knowledge of signal properties that mitigates the uncertainty problem. We evaluate the performance of this method by computing the probability of detecting signal occupancy in terms of the number of samples and the SNR of randomly generated signals. The results show a reliable detection even in low SNR and small number of samples.
\end{abstract} 
\section{Introduction}
\label{sec:intro}
Spectrum sensing is an important function to enable cognitive radios to detect the underutilized spectrum licensed to the primary systems and improve the overall spectrum efficiency \cite{CRN}. Some well-known spectrum sensing techniques are energy detection, matched filter and cyclostationary feature detection that have been proposed for narrowband sensing. In these methods, based on the signal properties, a decision is made to detect presence or absence of a primary user in the considered band \cite{CRN}.

Future cognitive radios should be capable of scanning a wideband of frequencies, in the order of few GHz \cite{CWBSSG}. In the wideband regime, the radio front-end can employ a bank of band-pass filters to select a frequency band and then exploit the existing techniques for each narrowband, but this method requires a large number of RF components \cite{ZTCS}.

Alternatively, in order to identify the locations of vacant frequency bands, the entire wideband is modelled as a train of consecutive frequency sub-bands \cite{CRN} and the total wideband is sampled classically or in a compressed way. After estimation of the spectrum from the obtained samples, the conventional spectrum sensing methods such as energy detection would be applied to detect the signal in each band. Classical sampling of a wideband  signal needs high sampling rate ADCs, which have to operate at or above the Nyquist rate. Clearly, this is a major implementation challenge. Recent work based on compressive sampling has been proposed to overcome the problem of high sampling rates in \cite{CWBSSG},\cite{ZTCS}. However, estimating the spectrum of a signal from its compressed samples is achieved by solving an optimization problem \cite{CWBSSG}, which is not an easy task. In addition, detection of the signal in each channel from its estimated spectrum needs knowledge of the signal and noise power that is not reliable because of uncertainty. 

By using the fact that the wireless signals in open-spectrum networks are typically sparse in the frequency domain, we propose a wideband spectrum sensing method that would bring substantial saving in terms of the sampling rate. The frequency band of interest is divided into a finite number of spectral bands, and the presence and absence of the signal in each spectral band is examined by considering the correlation matrix of the sampled data. In our method, estimation of the signal spectrum is skipped, and we directly detect the occupied channels from the sampled data in the time domain. Additionally, the problem of noise uncertainty is not of concern in this method, since the knowledge of signal and noise power is not needed. We evaluate this method by computing the probability of detecting signal occupancy in terms of the number of samples and signal to noise ratio (SNR). The outline of the paper is as follows:
The next section states the signal model and problem formulation. Section \ref{sec:model} introduces the spectrum sensing method and explains the functionality of each block in the model. In Section \ref{sec:results}, the simulation results and performance evaluation are presented and finally a conclusion is given in Section \ref{sec:con}.
  
\section{Problem statement}
\label{sec:problem}
The received signal $x(t)$ is assumed to be an analog wideband sparse spectrum signal, bandlimited to $[0,B_{max}]$. Denote the Fourier transform of $x(t)$ by $X(f)$. Depending on the application, the entire frequency band is segmented into $L$ narrowband channels, each of them with bandwidth $B$, such that $B_{max}=L\times B$. It is assumed that the signal bands are uncorrelated with each other. The channels  are indexed from $0$ to $L-1$. Those spectral bands which contain part of the signal spectrum are termed active channels, and the remaining bands are called vacant channels. Denote the number of such active channels by $N$. The indices of the $N$ active channels are collected into a vector 
\begin{equation}
\label{eq:chset}
\mathbf{b}=[b_1,b_2,\dots,b_N]
\end{equation}
which is referred to as the active channel set. 

\begin{figure}[t]
\psfrag {Spectrum} [][][.8]{PSD}
\psfrag {frequency[MHz]} [][][.8]{frequency[\mbox{MHz}]}
\centerline{\epsfig{figure=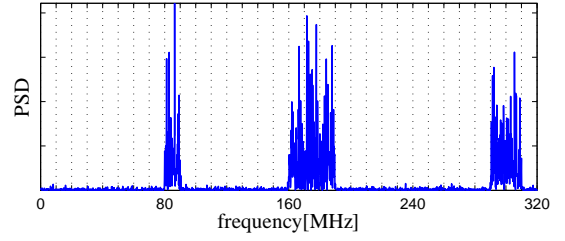},width=8.5cm}}
\caption{ Spectrum of a wideband signal received at the sensing radio with $L=32$ total bands and $N=6$ active channels. The active channel set is $\mathbf{b}=[8,16,17,18,29,30]$.}
\label{fig:sig}
\end{figure}

In the considered system, $N$ and $\mathbf{b}$ are unknown. 
However, we know the maximum channel occupancy which is defined as 
\begin{equation}
\label{eq:occ}
\Omega_{max}=\frac{N_{max}}{L}
\end{equation}
where $N_{max}\ge N$ is the maximum possible number of occupied channels. Figure ~\ref{fig:sig} depicts the spectrum of a multiband signal at the sensing radio, which contains $L=32$ channels, each with a bandwidth of $B=10$ MHz. The signal is present in $N=\nolinebreak[4]6$ channels, and the active channel set is $\mathbf{b}=[8,16,17,18,29,30]$.

The problem is, given $B_{max}$, $B$ and $\Omega_{max}$, to find the presence or absence of the signal in each spectral band or equivalently find the active channel set, $\mathbf{b}$, at a sub-Nyquist sample rate. 
\section{Wideband spectrum sensing model}
\label{sec:model}
\begin{figure*}[t]
\psfrag{xi[n]}{$x_i$}
\psfrag{x(t)}{$x(t)$}
\psfrag{Delay}{$z^{-c_i}$}
\psfrag{xi(n-c/L)}{$x_{d_i}$}
\psfrag{R=<x,x>}{$\frac{1}{M}\mathbf{x}_d\mathbf{x}_d^*$}
\psfrag{R}{$\mathbf{\hat{R}}$}
\psfrag{k}{${\mathbf{\hat{b}}}$}
\psfrag{Multicoset Sampler}{Multicoset Sampler}
\psfrag{Correlation matrix}{Sample Correlation matrix}
\psfrag{Subspace Methods}{Subspace Analysis}
\psfrag{fs=a fmax}{$f_{avg}=\alpha B_{max}$}

\centerline{\epsfig{figure=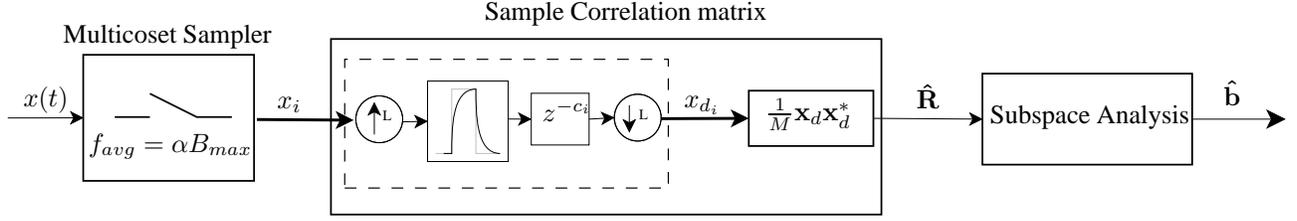},width=17cm}}
\caption{ Proposed wideband spectrum sensing model.}
\label{fig:model}
\end{figure*}

The proposed model for wideband spectrum sensing is illustrated in Figure ~\ref{fig:model}. The analog received signal at the sensing cognitive radio is sampled by the multicoset sampler at a sample rate lower than the Nyquist rate. The sampling reduction ratio is affected by the channel occupancy and multicoset sampling parameters. The outputs of the multicoset sampler are partially shifted using a multirate system, which contains the interpolation, delaying and downsampling stages. Next, the sample correlation matrix is computed from the finite number of obtained data. Finally, the correlation matrix is investigated to discover the position of the active channels by subspace methods. In this section each block of the model is described in detail.

\subsection{Multicoset sampler}
\label{subsec:sampler}
The analog wideband signal $x(t)$ is sampled using a multicoset sampling scheme introduced in \cite{OptBres}. The multicoset sampler provides $p$ data sequences for $i=1,...,p $, given by
\begin{equation}
 x_i(m)=x[(mL+c_i)/B_{max}],	m\in \mathbb{Z},
\end{equation}
where $\{c_i\}$, is a randomly selecting $p$ numbers out of the set $\L =\{0,1,...,L-1\}$ \cite{OptBres},\cite{BEldar}. 

The average sample rate of this scheme is $f_{avg} = \alpha B_{max}$, \cite{OptBres} where
$
\label{eq:alpha}
\alpha=(\frac{p}{L})
$
is termed the sub-Nyquist factor. According to Landau's lower bound \cite{Landau}, $\alpha$ is lower bounded to the maximum channel occupancy, $\alpha\ge \Omega_{max}$.

The number of data sequences, $p$, should be chosen greater than the maximum number of active channels $N_{max}$, to satisfy the Landau's lower bound and provide enough equations to find the unknown parameters.

\subsection{Sample correlation matrix}
\label{subsec:corr}
The main purpose of this section is to relate the problem of spectrum sensing with the problem of parameter estimation. Towards this goal, the correlation matrix of a special configuration of sampled data is computed. In order to achieve this, the following configurations are applied on the sampled data. 

First, each $x_i(m)$ sequence is over-sampled by a factor $L$, such that 
$$
x_{u_i}[n]= \left\{\begin{array}{rcc}
x_i(\frac{n}{L}), &  n=mL, m\in \mathbb{Z}   \\ 0, & \mbox{otherwise}
\end{array}\right.
$$
and then it is filtered to obtain 
$
x_{h_i}[n]=x_{u_i}[n]* h[n],
$
where $h[n]$ is the interpolation filter with the frequency response of 
\begin{equation}
\label{eq:Hf}
H(f) = 
\begin{cases} 
1,  & f\in [0,B] \\
0, & \mbox{otherwise}. 
\end{cases}
\end{equation}
Next, the output filtered sequence is delayed with $c_i$ samples such that
\begin{equation}
\label{eq:xci}
x_{c_i}[n]=x_{h_i}[n-c_i].
\end{equation}

Let us define $\mathbf{y}(f)$ as the known vector of observations 
\begin{equation}
\label{eq:y}
\mathbf{y}(f)=
\begin{bmatrix}
X_1(f),
X_2(f),
\dots ,
X_p(f)
\end{bmatrix}^T
\end{equation}
where the superscript $T$ denotes the transpose, and $X_i(f)$ is the $DFT$ of the sequence $x_{c_i}[n]$. Also, $\mathbf{x}(f)$, the unknown vector of the signal spectrum parameters is defined as 
\begin{equation}
\label{eq:z}
\mathbf{x}(f)=
\begin{bmatrix}
{X}(f+b_1B)\\
{X}(f+b_2B)\\
\vdots \\
{X}(f+b_NB)\\
\end{bmatrix},
\quad f \in [0,B]
\end{equation}
where $X(f+b_iB),f\in [0,B]$, are the frequency elements of the signal in the active band indexed by $b_i$. 

After applying Fourier transform on both sides of (\ref{eq:xci}) and expressing the result in matrix form, the data model in the frequency domain is given by \cite{BEldar}
\begin{equation}
\label{eq:yaz}
\mathbf{y}(f)= \mathbf{A(b)} \mathbf{x}(f)+\mathbf{n}(f),\quad f \in [0,B]
\end{equation}
where $\mathbf{A(b)}\in \mathbb{C}^{p\times N}$ is the modulation matrix given by \cite{BEldar}
\begin{equation}
\label{eq:Ak}
\mathbf{A(b)}(i,k)= B \exp{\left(\frac{j2\pi c_i b_k}{L}\right)}
\end{equation}
and $\mathbf{n}(f)$ is the frequency representation of the noise. For simplicity, we assume that $\mathbf{n}(f)$ is a Gaussian complex noise with distribution of $\mathcal{N}(0,\sigma^2 \mathbf{I})$, which is also uncorrelated with the signal. 

The model in (\ref{eq:yaz}) is a classical signal model that relates the observation vector, $\mathbf{y}(f)$, with the unknown signal spectrum vector, $\mathbf{x}(f)$, via the modulation matrix $\mathbf{A(b)}$. Note that the unknown signal parameter $\mathbf{b}$ is also the active channel set that is desired in the problem of spectrum sensing. Therefore, with this configuration, the problem of wideband spectrum sensing is turned into the problem of finding the model parameter $\mathbf{b}$ with minimum length $N$, subject to the data model (\ref{eq:yaz}). This is a combined detection-estimation problem, where we want to estimate the number as well as the parameters of the signals. Due to the computational complexity involved, the problem is solved in two steps: first the number of parameters is detected, and then, with an estimate of the number of parameters $\hat{N}$ at hand, the parameters of the signal are estimated. An approach based on the correlation matrix of the observations is employed here for solving this problem.

The correlation matrix of observation vector is defined as 
\begin{equation}
\label{eq:corr}
\mathbf{R}= E[\mathbf{y}(f)\mathbf{y}^*(f)]= \mathbf{A(b)} \mathbf{P} \mathbf{A^*(b)}+ \sigma^2 \mathbf{I}
\end{equation}
where $( )^*$ denotes the Hermitian transpose, and 
\begin{equation}
\label{eq:Z}
\mathbf{P}= E[\mathbf{x}(f) \mathbf{x}^*(f)]
\end{equation}
is the correlation matrix of the signal vector \cite{DOAM}. 

Since the distribution of the signal is unknown the real correlation matrix $\mathbf{R}$ cannot be achieved. Hence, we estimate $\mathbf{R}$ from the integration of $[\mathbf{y}(f) \mathbf{y}^*(f)]$ over the interval  $[0,B]$. However, from Parseval's identity it can be computed directly in the time domain from the  sequences $x_{c_i}[n]$ at the sample rate of $B_{max}$. Since each $x_{c_i}[n]$ sequence is the output of a narrowband filter, the reduced bandwidth output signal can be easily accommodated within a lower output sample rate. This means that the computations do not need to be performed at the high sample rate, $B_{max}$. Thus, the sequences are down-sampled by $L$, the reduced bandwidth factor, such that 
\begin{equation}
x_{d_i}(m)=x_{c_i}[mL]
\end{equation}

The total process of oversampling, filtering, delaying and downsampling from $x_i(m)$ to $x_{d_i}(m)$ is viewed as a fractional shifting of the sequence $x_i(n)$. Defining the snapshot vector $\mathbf{x}_d(m)$ as 
$$
\label{eq:xd}
\mathbf{x}_d(m)=
\begin{bmatrix}
x_{d_1}(m) \\
x_{d_2}(m) \\
\vdots \\
x_{d_p}(m)
\end{bmatrix},
$$ 
the $p\times p$ sample correlation matrix from $M$ samples of the partially shifted sequence is computed from the formula \cite{DOAM}
\begin{equation}
\label{eq:scorr}
\mathbf{\hat{R}}=\frac{1}{M} \sum_{m=1}^M \mathbf{x}_d(m) \mathbf{x}_d^*(m).
\end{equation}
Under suitable assumptions $\mathbf{\hat{R}} \rightarrow \mathbf{R}$ when $M\rightarrow \infty$.

\subsection{Subspace Analysis}
\label{sec:subspace}
The signal in each spectral band is assumed to be uncorrelated with the other bands. Hence, the correlation matrix of the signal vector, $\mathbf{P}$, is always full-rank and the class of subspace methods is applicable which is described in this section.

\subsubsection{Estimating the Number of Active channels}
\label{subsec:decomp}
In a geometrical view of (\ref{eq:corr}), any orthogonal vector to $\mathbf{A(b)}$ is an eigenvector of $\mathbf{R}$ with corresponding eigenvalue $\sigma^2$. The remaining eigenvectors are all in the range space of $\mathbf{A(b)}$, and therefore are termed signal eigenvectors. The eigen-decomposition of $\mathbf{R}$ is partitioned into a signal and a noise subspace as \cite{DOAM}
\begin{equation}
\label{eq:decom}
\mathbf{R}=\mathbf{E}_s \mathbf{\Lambda}_s \mathbf{E}_s^*+\mathbf{E}_n \mathbf{\Lambda}_n \mathbf{E}_n^*
\end{equation}
where $\mathbf{\Lambda}_s$ and $\mathbf{\Lambda}_n$ are diagonal matrices of signal and noise eigenvalues respectively, and $\mathbf{E}_s$ and $\mathbf{E}_n$ are the matrices of the corresponding eigenvectors. The signal eigenvectors in $\mathbf{E}_s$ span the range space of $\mathbf{A(b)}$ , which is termed the signal subspace. For the noise eigenvector we have instead, $\mathbf{E}_n \perp \mathbf{A(b)}$ \cite{DOAM}. So, if the dimension of noise subspace is obtained, the orthogonality property can be exploited to find the signal parameters.

If we denote the ordered eigenvalues of $\mathbf{\hat{R}}$ by 
$$\lambda_1 \ge \lambda_2\ge \dots \ge \lambda_p$$ 
then it follows that in case of large enough number of samples, $\mathbf{\hat{R}} \rightarrow \mathbf{R}$ and the smallest $(p-N)$ eigenvalues of $\mathbf{\hat{R}}$ are all equal to $\sigma^2$ \cite{ITC}. Thus, for large $M$, the dimension of the signal vector can be determined from the multiplicity of the smallest eigenvalues of $\mathbf{\hat{R}}$. 

\begin{figure}[t]
\psfrag {Eigenvalues,[dB]} [][][.8]{$\quad \lambda_i,[\mbox{dB}]$}
\psfrag {index,i} {$i$}
\centerline{\epsfig{figure=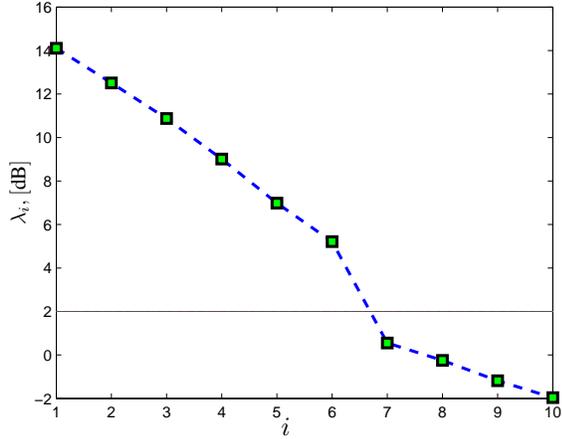},width=8.5cm}}
\caption{ Ordered eigenvalues of a sample correlation matrix with $M=61$. There are six eigenvalues greater than the threshold level.}
\label{fig:eigen}
\end{figure}

In realistic scenarios, the number of samples depends on the duration of the sensing period, which should be as short as possible, especially in case of time-varying channels. Hence the sample matrix $\mathbf{\hat{R}} \ne \mathbf{R}$. In the spectrum of ordered eigenvalues, the signal eigenvalues are still identified as the $N$ largest ones. But, the noise eigenvalues are no longer equal to each other \cite{EFT} and the separation between the signal and noise eigenvalues needs a threshold level. Depending on the noise power and the number of samples, the threshold level should be changed. Figure ~\ref{fig:eigen} depicts a typical set of ordered eigenvalues that are separated by a horizontal line as the threshold level. As one can see there are six eigenvalues greater than the threshold level. That specifies the number of active channels. 

To avoid the threshold setting for different tests, we applied information theoretic criteria for model order selection such as minimum description length (MDL). 
The number of active channels using the MDL criterion for $0\leq r \leq N_{max} $ is given by \cite{ITC},\cite{EFT}
\begin{equation}
\label{eq:mdl}
\hat{N}= \arg\min\limits_r -M(p-r) \log \frac{g(r)}{a(r)} +\frac{1}{2} r (2p-r) \log M 
\end{equation}
where $M$ is the number of samples in each sequence, $g(r)$ and $a(r)$ are the geometric and arithmetic mean of the $(p-r)$ eigenvalues of the correlation matrix respectively.

The probability of correct detection of $N$, namely, $Pr(\hat{N}=N)$, depends on the number of samples $(M)$, SNR and sub-Nyquist factor $(\alpha)$. Since evaluation of this detection algorithm in the Neyman-Pearson sense of finding the most powerful test that does not exceed a threshold probability of false alarm is computationally very expensive \cite{ITC}, we will evaluate the performance of this method numerically in Section \ref{sec:results}.

\begin{figure}[t]
\psfrag {|X(f)|} [][][.8]{$|X(f)|$}
\psfrag {P} [][][.8]{$P$}
\psfrag {MU} [][][.8]{$MU$}
\psfrag {(h)} [][][.8]{$(k)$}
\psfrag {frequency[MHz]} [][][.8]{frequency[\mbox{MHz}]}
\psfrag {h, channel index} [][][.8] {$k,\mbox{channel index}$}
\centerline{\epsfig{figure=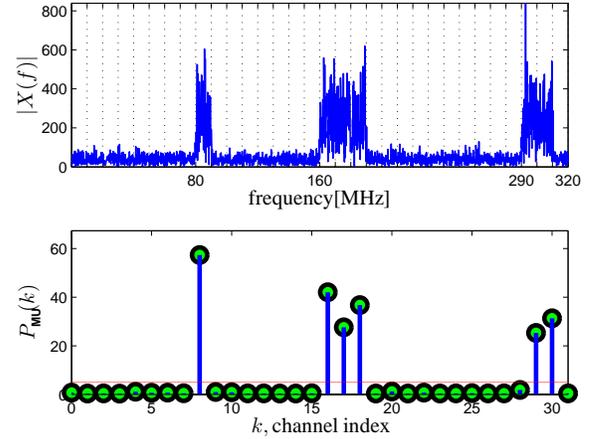},width=8.5cm}}
\caption{ Frequency representation, $X(f)$, and the corresponding $P_{MU}$ values of a typical wideband signal with $L=32$ channels. The position of six significant values specify the occupied channels $\mathbf{\hat{b}}=[8,16,17,18,29,30]$.}
\label{fig:music}
\end{figure}

\subsubsection{Active channel set recovery}
\label{subsec:chset}
After estimating the number of active channels, the $(p-\hat{N})$ smallest eigenvalues are specified as the noise eigenvalues. Denote the corresponding noise eigenvectors with $\mathbf{\hat{E}}_n$ as a $p\times (p-\hat{N})$ matrix. The location of the active channels can be recovered according to a MUSIC-Like algorithm as \cite{MUSIC}
\begin{equation}
\label{eq:music}
\providecommand{\norm}[1]{\lVert#1\rVert}
P_{MU}(k)=\frac{1}{\norm {\mathbf{a}_k \hat{\mathbf{E}}_n}^2}, \quad 0\leq k\leq L-1
\end{equation}
where
$
\providecommand{\norm}[1]{\lVert#1\rVert}
\norm{.}
$
denotes the 2-norm, $k$ is the channel index and $\mathbf{a}_k$ is a column of $\mathbf{A(b)}$, given by
\begin{equation}
\label{eq:ah}
\mathbf{a}_k= 
\begin{bmatrix}
e^{\frac{j2\pi k c_1 }{L}},
e^{\frac{j2\pi k c_2 }{L}},
\dots ,
e^{\frac{j2\pi k c_p }{L}}
\end{bmatrix}^T
\end{equation}

The algorithm generates $L$ values corresponding to the $L$ channels. If $k$ is the index of an active channel, $P_{MU}(k)$ is significant in that point, otherwise it will be a small value. The estimated active channel set, $\mathbf{\hat{b}}$, is determined by selecting the position of these significant values. Figure ~\ref{fig:music} depicts the frequency representation of the received signal, $X(f)$, and the corresponding $P_{MU}$ values of a typical wideband system. As illustrated in the figure, several significant values correspond to active channels are appeared where their locations specify the estimated active channel set. The other channels are interpreted as the vacant channels and can be used by the cognitive system to transmit.

\begin{figure}[t]
\psfrag {Pd} [][][.8]{$Pr(\hat{N}=6)$}
\psfrag {M} [][][.8]{$M$}
\centerline{\epsfig{figure=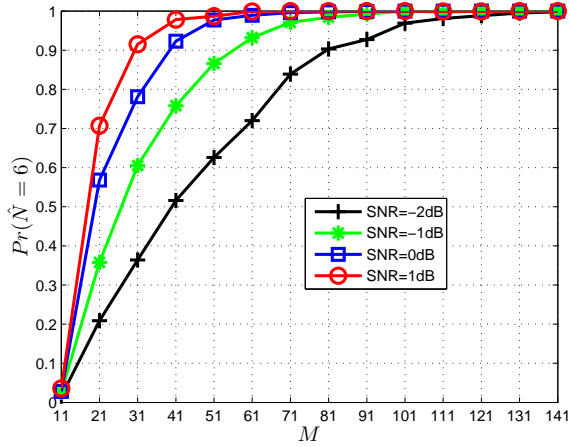},width=8.5cm}}
\caption{ $Pr(\hat{N}=6)$, probability of detecting number of active channels versus $M$ and SNR for the simulated wideband system.}
\label{fig:pdq}
\end{figure}

\begin{figure}[t]
\psfrag {Pd}[][][.8]{$P_{\text{d}}$}
\psfrag {Pf}[][][.8]{$P_{\text{f}}$}
\psfrag {M} [][][.8]{$M$}
\centerline{\epsfig{figure=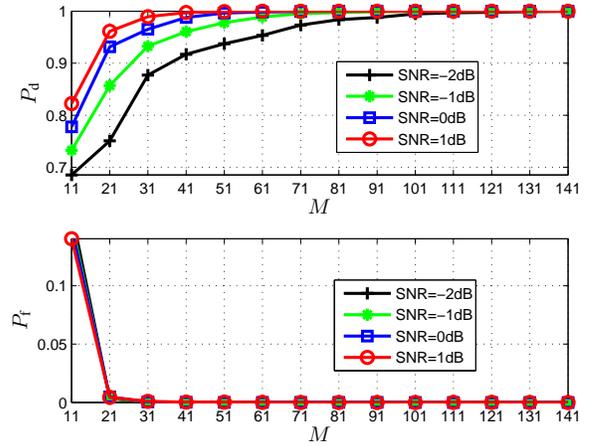},width=8.5cm}}
\caption{ Detection performance of the proposed model versus $M$ and SNR for the simulated wideband system..}
\label{fig:pdfk}
\end{figure}

\section{Numerical results}
\label{sec:results}
In this section, we illustrate the performance of our method using Monte Carlo simulations.

The received signal at the cognitive radio sensing is generated from the model
$$
x[n]= \sum_{i=1}^{N} (r_i[n]*h[n]) \exp(j2\pi f_i n /B_{max})+ w[n]
$$
where $*$ shows a convolution between $r_i[n]\sim \mathcal{N}(0,\sigma_i^2)$ and $h[n]$, the low pass filter defined in (\ref{eq:Hf}). The output of the convolution is placed at the carrier frequency $f_i$, and corrupted by $w[n] \sim\mathcal{N}(0,1)$, the additive white Gaussian noise. 

The wideband of interest is in the range of $[0,320]$\nolinebreak[4] MHz, containing $32$ channels of equal bandwidth of $B=10$MHz. The signal variance is chosen such that the received SNR of all active channels are equal. Figure ~\ref{fig:sig} depicts the spectrum of the signal model with $N=6$ active bands located at different unknown carriers. Given $B_{max}=320$MHz, $\Omega_{max}= 0.25$ and $B=10$MHz, it is desired to find the positions of occupied and vacant channels at a sub-Nyquist sampling rate. 

A multicoset sampler with parameters $L=32, p=10$ is used to sample the signal at the average sample rate of  $f_{avg}=\nolinebreak[4]100$ MHz, which is $\alpha \approx 0.3$ of the Nyquist rate. Ten $c_i$ numbers are selected randomly out of the set $\L$.

We first investigate the performance of estimating $N$. Using 1000 Monte Carlo simulations, and various values of $M$ and SNR, we obtain an estimate of the number of active channels from the eigenvalues of the sample correlation matrix using (\ref{eq:mdl}). For a choice of SNR and different values of $M$, we compute the empirical probability of detecting six active channels. As $M$ and SNR are increased, we expect the estimator to detect the correct value with high probability. 

Figure \ref{fig:pdq} depicts the computed $Pr(\hat{N}=6)$ for different number of samples and SNR. It is seen at SNR=1dB after $M\ge 41$, the estimator is able to detect the correct value with high probability. However, in lower SNR, more samples are needed to achieve a high probability. 

Next, we evaluate the detection performance of the proposed method by computing the probability of detecting the signal occupancy as 
\begin{equation}
\label{eq:pdk}
P_{\mbox{d}}=\frac{1}{N} \sum_{i=1}^N Pr(b_i\in \mathbf{\hat{b}}|b_i\in \mathbf{b})
\end{equation}  
and the false alarm probability as 
\begin{equation}
\label{eq:pfk}
P_{\mbox{f}}=\frac{1}{L-N} \sum_{i=1}^{L-N} Pr(b_i^c\in \mathbf{\hat{b}}|b_i^c\in \mathbf{b}^c)
\end{equation}
where $\mathbf{b}^c=\L-\mathbf{b}$ is the complement set of $\mathbf{b}$. 

The $P_{\mbox{d}}$ and $P_{\mbox{f}}$ are computed from (\ref{eq:pdk}) and (\ref{eq:pfk}) for different values of $M$ and SNR and the results are shown in Figure ~\ref{fig:pdfk}. The results exhibit outstanding detection performance even in low SNR and small $M$. It is seen that at SNR=1dB, after $M\ge41$, the proposed model detects the occupied channels with probability close to one. The value of $P_{\mbox{f}}$ decreases dramatically with increasing $M$, such that after $M\ge 21$ it becomes almost zero for all SNR values. Comparing Figure~\ref{fig:pdq} and ~\ref{fig:pdfk} reveals that with correct estimation of $N$, the perfect detection of channel occupancy is possible. In the other words, the accuracy of the proposed model only depends on the MDL estimator and would be improved using other techniques such as the method in \cite{EFT}.

\section{Conclusion}
\label{sec:con}
A method of wideband spectrum sensing for cognitive radio is proposed to mitigate the limitations of high sampling rate, high complexity and noise uncertainty. The proposed technique utilizes a multicoset sampling scheme that can use arbitrarily low sampling rates close to the channel occupancy. With low spectrum utilization assumption, this would bring substantial savings in terms of the sampling rate. The coset samples are fractional shifted and used to compute the correlation matrix of the signal. The computation cost of this step is linear in the amount of data. The problem of spectrum sensing is turned into the problem of parameter estimation and then is solved by subspace methods that today are standard tools in signal processing. We evaluate the detection performance of this method for a typical case. The results show that even in low SNR with taking enough number of samples a perfect detection is possible. For a typical wideband system with $\Omega_{max}=0.25$ and SNR=1dB by taking $M=31$ samples, at $\alpha\approx 0.3$ of the Nyquist rate, $P_{\mbox{d}}=0.99$ and $P_{\mbox{f}}=10^{-3}$ are achieved.
\bibliographystyle{IEEEbib}
\bibliography{strings}

\end{document}